\documentclass[printer]{aa}
\usepackage{graphicx}
\usepackage{natbib}
\bibpunct{(}{)}{;}{a}{}{,} 
\usepackage{txfonts}
\titlerunning{Physical and Kinematical properties of APM 08279+5255}

\def\kms{km ${\rm s}^{-1}$}
\def\sn1{${\rm s}^{-1}$}
\def\cmd{${\rm cm}^{-3}$}
\def\cmn{${\rm cm}^{-2}$}
\def\ergs{ergs ${\rm s}^{-1}$}
\def\flux{ergs cm$^{-2}$ ${\rm s}^{-1}$}
\def\xmm{{\it XMM-Newton}}
\def\chan{{\it Chandra}}
\def\apm{APM~08279+5255}
\def\mr{MR~2251$-$178}
\def\sp{\space}

\begin{document}
\bibliographystyle{aa}

   \title{Physical and kinematical properties of the X-ray absorber in the broad absorption line quasar APM 08279+5255}

\author{
J.M.~Ram{\'i}rez}   
\offprints{Ram{\'i}rez}

\institute{Max-Planck-Institut f\"{u}r extraterrestrische Physik, D-85741 Garching, Germany}


   \date{}

  \abstract
  {We have re-analyzed the X-ray spectra of the \emph{gravitational lensed}
high-redshift BAL QSO \apm, observed with the \xmm \sp and \chan \sp observatories.
Previous studies (Hasinger et al. 2002; Chartas et al. 2002) detected
unusual, highly-ionized iron absorption features, but differed in their
interpretation of these features, regarding the kinematical and
ionization structure.}{We seek one physical model that can be successfully
applied to both observations.}{For the first time we have performed detailed photoionization
modeling on the X-ray spectrum of \apm.}{The absorbing gas
in \apm \sp can be represented by a two-absorbers model with
column densities $N_H(1)\approx 7\times 10^{22}$ \cmn, $N_H(2)\approx 6\times 10^{22}$ \cmn,
and ionization parameters $\log\xi(1)\approx 1.5$ and $\log\xi(2)\approx 3$,
with one of them (the high-ionization component) outflowing at $v\approx 0.18(\pm 0.01)c$,
carrying large amount of gas out of the system.
We find that the \chan \sp spectrum of \apm \sp requires the same Fe/O ratio overabundance
(previously) indicated by the \xmm \sp observation, showing that both absorber components underwent
similar chemical evolution and/or have similar origin.}{}


   \maketitle
%

\section{Introduction}

Broad absorption line (BAL) quasi-stellar objects (QSOs) are 
objects displaying in their spectra broad (FWHM $\approx$ 10\,000
\kms) absorption lines in the rest-frame ultraviolet (UV),
originated in outflows of matter from the central engine of
QSOs \citep{foltz1990a,weymann1991a}. The outflow velocity
may reach up to $0.2c$ \citep[e.g.,][]{foltz1983a}. Determining the
relationship between the material absorbing the X-rays and
the one absorbing the UV radiation is
key to our understanding of the geometry and the physical state of
the medium surrounding the vicinity of supermassive black holes
\citep[e.g.,][]{mathur1995a,murray1995a,hamann1998a,proga2000a}.

Before \chan/\xmm \sp missions, detections of BAL quasars in X-ray were rare.
Usually, these object are X-ray weak \citep[e.g.,][]{gallagher2006a},
sometimes interpreted as strong excess absorption.
\chan \sp and \xmm \sp observations of BAL QSOs, have provided new
constraints in the amount of absorption toward selected objects
\citep[e.g.,][]{sabra2001a,oshima2001a,gallagher2002a,gallagher2006a},
indicating large column densities $N_H \gtrsim 10^{23}$ \cmn.

The BAL QSO \apm \space at redshift $z=3.91$ \citep{irwin1998a} is one
of the most luminous objects in the universe, further magnified
by gravitational lensing by a factor of $\sim 50-100$ \citep[e.g.,][]{ledoux1998a}.
It was detected with the Submilimiter Common-User Bolometric Array,
implying an apparent far-infrared luminosity greater than
5$\times 10^{15} L_{\sun}$ \citep{lewis1998a}. The optical
spectrum, obtained with the High Resolution Echelle Spectrometer
at the Keck telescope \citep{ellison1999a}, along with a detailed
study of the physical conditions in the BAL flow of the QSO
by \cite{srianand2000a}, allowed them to conclude that the corresponding
gas stream, outflowing with velocities of up to 12\,000 \kms, is
heavily structured and highly ionized.

The quasar \apm \space was observed twice with \xmm \sp \citep[][ hereafter H02]{hasinger2002a}.
In both observations the quasar is observed clearly out to 12 keV, which
corresponds to almost 60 keV in the rest frame.
The most apparent feature in the \xmm \sp spectrum is an absorption-like
feature around 1.55 keV (which they interpret as an absorption edge), 
corresponding to $\sim 7.7$ keV in the rest frame
of \apm. The high-inferred iron abundance at the high redshift, corresponding to a
young age of the universe, is of great interest in the context of chemical
enrichment models, and provides constraints on the early star formation
history of the universe and on its cosmological parameters
\citep[e.g.,][]{hamann1993a,hasinger2002a,komossa2003a}.

The quasar \apm \space was also observed with \chan \sp \citep[][ hereafter C02]{chartas2002a}.
The \chan \sp spectrum shows a similar absorption feature as the \xmm \sp
observation, but the feature led to a different interpretation.
In particular, C02 modeled the spectrum with two
absorption lines at 8.1 keV and 9.8 keV in the rest frame of the quasar,
interpreted as \ion{Fe}{xxv} K lines. If the \chan \sp
interpretation of the data is right, the {\it bulk} velocity
of the X-ray BALs is $\sim 0.2c-0.4c$.
The presence of similar outflow velocities has been claimed in a
few other AGN X-ray spectra \citep[e.g.,][]{pounds2003a,chartas2003a,pounds2006a},
but alternative interpretations of the same spectra have been
proposed, which do not require these relativistic outflow velocities
\citep[e.g.,][]{kaspi2006a}.

New models \citep[e.g.,][]{elvis2000a,proga2000a,proga2007a} 
predict that a large fraction of the accreted
matter into the region of the compact object, is expelled out
again in the form of high-velocity outflows. Broad UV
absorption features, with velocities $\sim$ 0.05$c-0.1c$, are associated
with this material
through acceleration mechanisms like acceleration by gas
pressure \citep[e.g.,][]{weymann1982a,begelman1991a} due to dust
\citep[e.g.,][]{voit1993a,yun1995a} and acceleration due to radiative
pressure by spectral lines \citep[e.g.,][]{drew1984a,shlosman1985a,arav1994a,dekool1995a,murray1995a,proga2000a}
observed in about 10\% of luminous high-$z$ quasars
\citep{laor2002a}, implying
that these outflows are an important component of the general picture
of AGNs.
Furthermore, these models also predict that
in order for the UV material to reach such high velocities a \emph{shield}
made by a high column density of gas ($N_H\sim 10^{22}-10^{24}$ \cmn) 
must absorb radiation in
the X-ray band, in this manner preventing the destruction of the UV material.
The new generation of X-ray observatories, \xmm \sp and \chan \sp
give us an unique opportunity to study in detail the X-ray component
of the BAL QSOs, contributing in our knowledge about the dynamical
and physical evolution of the center of these systems, and in the
case of high-redshited BAL QSOs, the gas enrichment history in the
early universe \citep{hamann2004a}.

However, none of the previous studies include a self-consistent photoionization
modeling of the X-ray spectrum of \apm. Any constraint on the
ionization level of the absorbing gas and its connection with the
kinematical properties of the BAL outflow is of major interest
to elucidate some of the greatest discrepancies between
these two proposed models. Furthermore, a model that
can separate the differences between observations (\xmm \sp {\it vs} \chan),
or unify both in a single frame, is highly desirable.
We report a spectral analysis (made of these two observations
separated by $\sim 2$ weeks in the rest frame)
of the high-redshifted BAL QSO \apm, and present one model that
might reconcile both observations in the same physical context,
shedding new light on the ionization degree, kinematics and evolution
of this system, in connection with its cosmological consequences.
We use the following cosmological parameters:
H$_0=70$ \kms \space Mpc$^{-1}$, $\Omega_M=0.3$ and $\Omega_{\lambda}=0.7$.

\section{X-ray observations and Spectral analysis}

The quasar \apm \space was observed twice with \xmm. The first
X-ray observation was made on 2001 October 30 with
$\sim 15$ ks of exposure time 
(referred as XMM1 in Table 2 of H02).
A significantly longer ($\sim 100$ ks) observation
was made in 2002 April 28$-$29
(referred as XMM2 in Table 2 of H02).
Taking advantage of the improvement
in calibration and effective areas used in \xmm \sp
observations, we re-processed the primary {\tt event} file
and extracted the spectrum, using the most updated
\xmm \sp Science Analysis System (SAS, version 7.0.0),
and the standard data pipeline processing.
In our analysis, we used the $\sim 88.8$ ks \chan \sp ACIS-I
observation of \apm \space (see Table 1 of C02),
and extracted the spectrum, built ancillary and redistribution
matrix files with CIAO\footnote{Chandra Interactive Analysis of Observations.
{\tt http://cxc.harvard.edu/ciao/}}
version 3.4. Details on calibration fluxes and count
rates can be found in the respective works.

We will present the spectral analysis in three steps:
1) By showing results from C02 (below in this section),
2) comparing them
with H02 (below in this section), and 3) we introduce our
own approach highlighting the differences between these sets
of data (\xmm \sp {\it vs} \chan, in Sect. \ref{var}).
Then we proceed to three more steps:
1) fitting the edge and line models; 2) fitting photoionization models
to the \xmm \sp and \chan \sp data separately; and 3) fitting jointly in the context
of one consistent model (in Sect. \ref{photo}).

In C02, the authors are able to fit an absorbed power-law with an intrinsic neutral absorber
in the frame of the quasar, to the \chan \sp spectrum of \apm. They find
a best-fit photon index $\Gamma\approx 1.7$ and the column density of the
absorber $N_H\approx 6 \times 10^{22}$ \cmn.
Significant residuals are found at $\sim 8$ keV (rest-frame). This is the source of
an important controversy. By the time C02 was under review, the authors had published H02,
in which a good description of the spectrum is proposed by fitting a model
that included an absorption edge 
({\tt zphabs $\times$ Pl $\times$ 1-Edge} or {\tt Edge} model) 
at $\sim 8$ keV to the \xmm \sp observation
of this object.
Despite the fact that \chan \sp data provides values for the parameters for
the {\tt Edge} model, its $\chi_{\nu} ^2$ is statistically poor
($\sim 1.36$ for 107 degrees of freedom [d.o.f]). Moreover, C02 tried to
model this feature with a {\tt Single-Gaussian} line model (the one at $\sim$ 8.1 keV)
and the fit gets a bit worse ($\chi_{\nu} ^2 \sim 1.41$ for 106 d.o.f).
They found a better fit with a {\tt Two-Gaussian} lines model
at $\sim 8.1$ keV and $\sim 9.8$ keV (rest frame), which indeed significatively improved
the description of the data over the {\tt Edge} model.
A visual comparison between the two sets of data suggests that the absorption
feature is present in both cases, but the line-shape of the feature appears to be
weakened in the \xmm \sp observation.

As noted by H02, a major improvement 
(in the description of the \xmm \sp spectrum of \apm)
is found by
adding an absorption edge to the absorbed power-law (Fit 2 in Table \ref{tbl1}).
Our best-fit model for the \xmm \sp 
data provides an edge energy at $\approx$ 7.7 keV (in the rest frame)
and an optical depth $\tau\approx0.3$. This was interpreted by the previous
authors as a Fe edge, compatible with ionization potentials of iron
from Fe~{\sc xv} to Fe~{\sc xviii}, implying a significant ionization
of iron. Indeed, in that work the authors compared the absorption 
seen in the spectrum
at low-energies and arrived at the conclusion
that an overabundance of iron ($\sim 2-5$) with respect to lower
$Z$ elements (O, Ne, Mg, Si and S) is necessary to account for
this feature ($N_{\rm Fe} \approx 1.5 \times 10^{19}$ \cmn).

In all our fits, we have included a neutral absorption
with the Galactic value fixed to $N_H=3.9 \times 10^{20}$
\cmn \space \citep{stark1992a}. We begin our spectral analysis by fitting
an absorbed power-law with an intrinsic neutral absorber
in the frame of the quasar ($z=3.91$). Our XMM (EPIC pn+MOS) and
XMM EPIC pn data \citep{struder2001a}, 
are consistent with each other and with the \chan \sp
fit (included also the H02 fits for comparison purposes).
The photon index is $\Gamma \sim 1.7-2$ and the column density of the
absorber is $N_H \sim 6 \times 10^{22}$ \cmn.
Table \ref{tbl1} quotes the parameter values of the model along
with the {\it goodness-of-fit} in terms of a global $\chi^2$ for
the d.o.f used for each dataset
(bins included in the range $0.2-10$ keV minus the number of free
parameters of the model), and the {\it global} $\chi^2-$probability
($P_{\nu,\chi^2}$), which gives the probability of exceeding $\chi^2$ for
$\nu$ degrees of freedom. Additionally, we introduce a criterion to assess
the fits close to the absorption feature, thus we take the global parameters
of the model and compute $\chi^2$ in the band $(1-2)$ keV (observed frame),
along with its corresponding {\it local} $\chi^2-$probability.

\begin{table*}
\caption{Fit Results and comparison between sets of data.
{\it XMM-Newton vs Chandra} observations of \apm.}
\label{tbl1}
\centering
\begin{tabular}{l c c c c c}
\hline \hline
Parameter&
H $^{\rm a}$&
C $^{\rm b}$&
(pn+MOS)$^{\rm c}$&
pn$^{\rm d}$&
Cn$^{\rm e}$
\\
\hline
\multicolumn{5}{l}{\small Fit 1: {\tt Pl} and Neutral Absorption at Source}\\ \hline

$\Gamma$ 			   & $2.04_{-0.03} ^{+0.03}$ & $1.72 _{-0.06} ^{+0.03}$ &  $1.99 _{-0.02} ^{+0.02}$ &  $2.05 _{-0.03} ^{+0.03}$		& $1.79 _{-0.03} ^{+0.03}$ \\
Norm$^{\rm f}$     		   & $1.30_{-0.05} ^{+0.05}$ &                          &  $1.23 _{-0.02} ^{+0.02}$ &  $1.25 _{-0.02} ^{+0.02}$		& $1.02 _{-0.02} ^{+0.02}$ \\
$N_H$ ($\times 10^{22}$ \cmn) 	   & $6.92_{-0.32} ^{+0.32}$ & $6.0  _{-0.8} ^{+0.8}  $ &  $5.95 _{-0.21} ^{+0.21}$ &  $6.28 _{-0.25} ^{+0.26}$		& $7.18 _{-0.38} ^{+0.41}$ \\	
$\chi^2$/(d.o.f) 		   & 481.8/(365) 	     & 182.1/(109)     	  	&  564.5/(478)     	    &  330/(290)  			& 229.7/(184) \\
$P_{{\rm global}}[{\rm Fit~1}]$    & $2 \times 10^{-5}$      & $1.5 \times 10^{-5}$     &  $3 \times 10^{-3}$       &  0.05				& 0.01 \\
$\chi^2(1-2~{\rm keV})$/(d.o.f)    & \ldots                  & 	\ldots  		&  134.7/(59)      	    &  72.4/(35)                      	& 85.7/(52)  \\
$P_{(1-2~{\rm keV})}[{\rm Fit~1}]$ & \ldots		     & 	\ldots			&  $2 \times 10^{-9}$	    &  $6 \times 10^{-5}$		& $2 \times 10^{-3}$ \\

\hline
\multicolumn{5}{l}{\small Fit 2: {\tt Pl} and Neutral Absorption at Source, and One Edge}\\ \hline

$\Gamma$            		    & $2.01_{-0.03} ^{+0.03}$ 	&                  		& $1.96 _{-0.02} ^{+0.02}$ & $2.02 _{-0.03} ^{+0.03}$ & $1.77 _{-0.02} ^{+0.02}$ \\
Norm$^{\rm f}$         		    & $1.37_{-0.06} ^{+0.06}$ 	&                  		& $1.28 _{-0.02} ^{+0.02}$ & $1.31 _{-0.02} ^{+0.02}$ & $1.09 _{-0.01} ^{+0.01}$ \\
$N_H$ ($\times 10^{22}$ \cmn) 	    & $7.34_{-0.34} ^{+0.34}$ 	&                  		& $6.20 _{-0.21} ^{+0.22}$ & $6.54 _{-0.26} ^{+0.27}$ & $7.50 _{-0.24} ^{+0.25}$ \\
$E_{\rm edge}$ (keV) 		    & $7.68_{-0.10} ^{+0.10}$ 	& $7.68 _{-0.25} ^{+0.21} $	& $7.70 _{-0.17} ^{+0.16}$ & $7.53 _{-0.19} ^{+0.21}$ & $7.70 _{-0.13} ^{+0.11}$ \\
$\tau$              		    & $0.46_{-0.05} ^{+0.05}$ 	& $0.37 _{-0.13} ^{+0.14} $	& $0.30 _{-0.06} ^{+0.06}$ & $0.33 _{-0.07} ^{+0.08}$ & $0.38 _{-0.05} ^{+0.05}$ \\
$\chi^2$/(d.o.f)  		    & 394.6/(362) 		& 146/(107)               	& 506/(476)                & 292/(288)		      & 190.1/(182) \\
$P_{{\rm global}}[{\rm Fit~2}]$     & 0.12      		& $6 \times 10^{-3}$          	& 0.17      		   & 0.44                     & 0.35 \\
$\chi^2(1-2~{\rm keV})$/(d.o.f)     & \ldots			& \ldots			& 74.5/(59)		   & 35.4/(35)		      & 53.7/(52) \\
$P_{(1-2~{\rm keV})}[{\rm Fit~2}]$  & \ldots                	& \ldots			& 0.09			   & 0.51		      & 0.46	  \\

\hline
\multicolumn{5}{l}{\small Fit 3: {\tt Pl} and Neutral Absorption at Source, and One Gaussian}\\ \hline

$\Gamma$              			&  \ldots & $1.73 _{-0.06} ^{+0.06}$ & $1.99 _{-0.02} ^{+0.02}$ & $2.04 _{-0.03} ^{+0.03}$ & $1.80 _{-0.02} ^{+0.02}$ \\
Norm$^{\rm f}$         			&  \ldots &                          & $1.25 _{-0.02} ^{+0.01}$ & $1.26 _{-0.02} ^{+0.02}$ & $1.06 _{-0.01} ^{+0.01}$ \\
$N_H$ ($\times 10^{22}$ \cmn) 		&  \ldots & $6.4  _{-0.9} ^{+0.8}  $ & $6.04 _{-0.20} ^{+0.23}$ & $6.28 _{-0.24} ^{+0.27}$ & $7.39 _{-0.24} ^{+0.25}$ \\
$E_{\rm line1}$       			&  \ldots & $8.05 _{-0.07} ^{+0.18}$ & $8.12 _{-0.03} ^{+0.04}$ & $7.97 _{-0.03} ^{+0.03}$ & $8.11 _{-0.01} ^{+0.01}$ \\
$\sigma_{\rm line1}$  			&  \ldots & \textless $0.140$        & $0.34 _{-0.11} ^{+0.13}$ & $0.15 _{-0.06} ^{+0.07}$ & $0.09 _{-0.02} ^{+0.02}$ \\
${\rm EW}_{\rm line1}$	(keV) 		&  \ldots & $0.23 _{-0.07} ^{+0.06}$ & $0.14 _{-0.11} ^{+0.18}$ & $0.16 _{-0.10} ^{+0.22}$ & $0.17 _{-0.04} ^{+0.04}$ \\
$\chi^2$/(d.o.f)			&  \ldots &  146.9/(106)             & 531.3/(476)              & 310.8/(288)		    & 198.4/(181) \\
$P_{{\rm global}}[{\rm Fit~3}]$     	&  \ldots &  $5.3 \times 10^{-5}$    & 0.04			& 0.18			    & 0.35	\\
$\chi^2(1-2~{\rm keV})$/(d.o.f) 	&  \ldots & \ldots		     & 98.8/(59)		& 53.2/(35)		    & 55.29/(52)  \\
$P_{(1-2~{\rm keV})}[{\rm Fit~3}]$   	&  \ldots & \ldots		     & $5 \times 10^{-4}$	& 0.03			    & 0.4	\\

\hline
\multicolumn{5}{l}{\small Fit 4: {\tt Pl} and Neutral Absorption at Source, and Two Gaussian}\\ \hline

$\Gamma$              			&  \ldots & $1.72 _{-0.05} ^{+0.06}$ & $1.98 _{-0.02} ^{+0.02}$ & $2.03 _{-0.03} ^{+0.03}$ & $1.79 _{-0.02} ^{+0.02}$ \\
Norm$^{\rm f}$         			&  \ldots &                          & $1.28 _{-0.02} ^{+0.02}$ & $1.29 _{-0.02} ^{+0.02}$ & $1.08 _{-0.01} ^{+0.02}$ \\
$N_H$ ($\times 10^{22}$ \cmn)  		&  \ldots & $6.7  _{-0.8} ^{+0.9}  $ & $6.20 _{-0.21} ^{+0.22}$ & $6.49 _{-0.25} ^{+0.27}$ & $7.47 _{-0.25} ^{+0.24}$ \\
$E_{\rm line1}$  (keV)     		&  \ldots & $8.05 _{-0.08} ^{+0.10}$ & $8.05 _{-0.02} ^{+0.02}$ & $7.93 _{-0.03} ^{+0.03}$ & $8.09 _{-0.01} ^{+0.01}$ \\
$\sigma_{\rm line1}$  (keV)		&  \ldots & \textless $0.140$        & $0.08 _{-0.03} ^{+0.02}$ & $0.09 _{-0.04} ^{+0.04}$ & $0.06 _{-0.01} ^{+0.01}$ \\
${\rm EW}_{\rm line1}$	(keV) 		&  \ldots & $0.24 _{-0.07} ^{+0.06}$ & $0.24 _{-0.07} ^{+0.12}$ & $0.30 _{-0.10} ^{+0.15}$ & $0.31 _{-0.10} ^{+0.19}$ \\
$E_{\rm line2}$       	(keV)		&  \ldots & $9.79 _{-0.19} ^{+0.20}$ & $9.53 _{-0.08} ^{+0.10}$ & $9.52 _{-0.09} ^{+0.10}$ & $9.72 _{-0.04} ^{+0.05}$ \\
$\sigma_{\rm line2}$  	(keV)		&  \ldots & $0.41 _{-0.16} ^{+0.19}$ & $0.80 _{-0.20} ^{+0.40}$ & $0.80 _{-0.10} ^{+0.30}$ & $0.70 _{-0.14} ^{+0.24}$ \\
${\rm EW}_{\rm line2}$	(keV)		&  \ldots & $0.43 _{-0.13} ^{+0.17}$ & $0.19 _{-0.08} ^{+0.10}$ & $0.21 _{-0.11} ^{+0.13}$ & $0.22 _{-0.09} ^{+0.14}$ \\
$\chi^2$/(d.o.f)			&  \ldots & 106.7/(103)              & 509.7/(474)      	&  293.2/(286)	           & 178.9/(179)\\
$P_{{\rm global}}[{\rm Fit~4}]$     	&  \ldots & 0.41		     & 0.13			& 0.39			   & 0.52 \\
$\chi^2(1-2~{\rm keV})$/(d.o.f) 	&  \ldots & \ldots           	     & 75.3/(59)                & 35.5/(35)                & 44.8/(52)  \\
$P_{(1-2~{\rm keV})}[{\rm Fit~4}]$    	&  \ldots & \ldots                   & 0.08       		& 0.51                     & 0.79       \\
\hline
\end{tabular}

\begin{flushleft}
All absorption-line and edge parameters are computed for the rest frame.
The error parameters are 90\% confidence limits.
(a) \cite{hasinger2002a}
(b) \cite{chartas2002a}
(c) The fits are carried out fitting simultaneously the pn+MOS data.
293 PI bins are from the pn data, and 188 bins are from the MOS data
(in the range 0.2$-$10 keV).
(d) Only pn data. (e) Our reprocessed \chan \sp data, with 187 bins.
(f) Power-law normalization, $\times 10^{-4}$ photons keV$^{-1}$cm$^{-2}$s$^{-1}$ at 1 keV
in the observed frame.
{\tt Pl}$\equiv$ Power-law.
\end{flushleft}

\end{table*}

We include in this analysis, fits of the \xmm \sp data of the {\tt Single} and 
{\tt Two-Gaussian} models, and the results are presented in Table \ref{tbl1},
for two sets of \xmm \sp data. One in which we fit simultaneously the
EPIC pn, MOS1 and MOS2 (pn+MOS) data and another only taking the EPIC pn data (pn). 
In the case of the
{\tt Single-Gaussian} line model, both sets are compatible with each other
and with the \chan \sp fit, and the three sets are in agreement with an
absorption feature with equivalent width (EW) $\sim 0.2$ keV.
For the {\tt Two-Gaussian} lines model, the best-fit parameters
lead to absorption lines at $\sim 8$ keV and $\sim 9.5$ keV (rest frame)
with EW $\sim 0.30$ keV and $\sim 0.20$ keV, respectively
(all of them compatible with the \chan \sp measurements, within the errors).
But, in this case, the $\chi^2-$probabilities are (for the \chan \sp fits) higher than those for
the {\tt Single-Gaussian} line model, and the {\tt Edge} model
(with 4 parameters more), in both senses, global and locally.
In principle, the fits of both models (Fit 2 and Fit 4) are formally acceptable
for both sets of data.
A more careful analysis, using physically-motivated arguments is required
to elucidate the ambiguity of this issue. A temporal variability study may help to solve some
aspect of the problem. We present a brief overview of the variability observed in the
X-ray spectrum of \apm, as seen from the point of view of the \xmm \sp and \chan
\sp data in the next section. A more quantitative physical discussion is given in
Sect. \ref{discuss}.
\section{Variability of the absorption structure at $\sim$ 7.7 keV $-$ 8 keV
\label{var}}
%
For this analysis, we worked on two sets of \xmm \sp data;     				%
a 16 ks exposure time observation of \apm \sp (XMM1), and a 100 ks 			%
exposure time observation (XMM2), see Table 1 of H02. 					%
As a first step, we were interested in searching for variability 			%
within each observation.								%
We looked at the count rate light curves for XMM1 and 					%
XMM2. We see slight changes in the light curve, although almost always 			%
within $\sim 10$\%, obtaining an average of $\sim 0.17$ cts~s$^{-1}$ for XMM1 		%
and $\sim 0.13$ cts~s$^{-1}$ for XMM2.							%
For the rest of the analysis we will only use the				%
data of the XMM2 observation of \apm, since it collects				%
one order of magnitude more photons with respect				%
to XMM1, increasing notably the signal-to-noise ratio,				%
and we refer to it as the \xmm \sp data.                                	%
\footnote{And from XMM2, we only make use of the EPIC pn data, since it has
a better calibration below 0.8 keV than the EPIC MOS
(see http://xmm.vilspa.esa.es/docs/documents/CAL-TN-0018-2-4.pdf). No significant
differences are seen between MOS and pn data, and the conclusions of the analysis
can be easily applied to the MOS data as well.}
We also looked for evidence of variability in the \chan \sp observation 	%
of \apm, but no significant deviation from the average $\sim 6.6\times 		%
10^{-2}$ cts~s$^{-1}$ was found, fully compatible with the count rate reported	%
by C02.								%
 							      %
                                                              %
                                                              %
                                                              %
                                                              %
                                                              %
%
 
Here, we discuss in some detail the differences we are seeing between the 	%
\xmm \sp and \chan \sp observations, which are separated 			%
by $\sim 2$ months in observed frame						%
($\sim 2$ weeks in the frame of the quasar), in the context of best-fit models.	%
To begin with, we will assume that indeed both sets of data are better 				%
represented by two different models ({\tt Edge} model {\it vs} {\tt 2-Absorption-lines} model),	%
and study how the \chan \sp data (better represented by the {\tt 2-Absorption-lines}		%
model, see C02 for details) behaves with the							%
best-fit model parameters extracted from fitting procedure, using \xmm \sp data			%
(specifically, the {\tt Edge} model parameters taken from Table 2 of H02), and			%
vicerversa (i.e., how the \xmm \sp data behaves with the best-fit model parameters 		%
found in fits using \chan \sp data).								%

\begin{figure}
\resizebox{\hsize}{!}{
\includegraphics{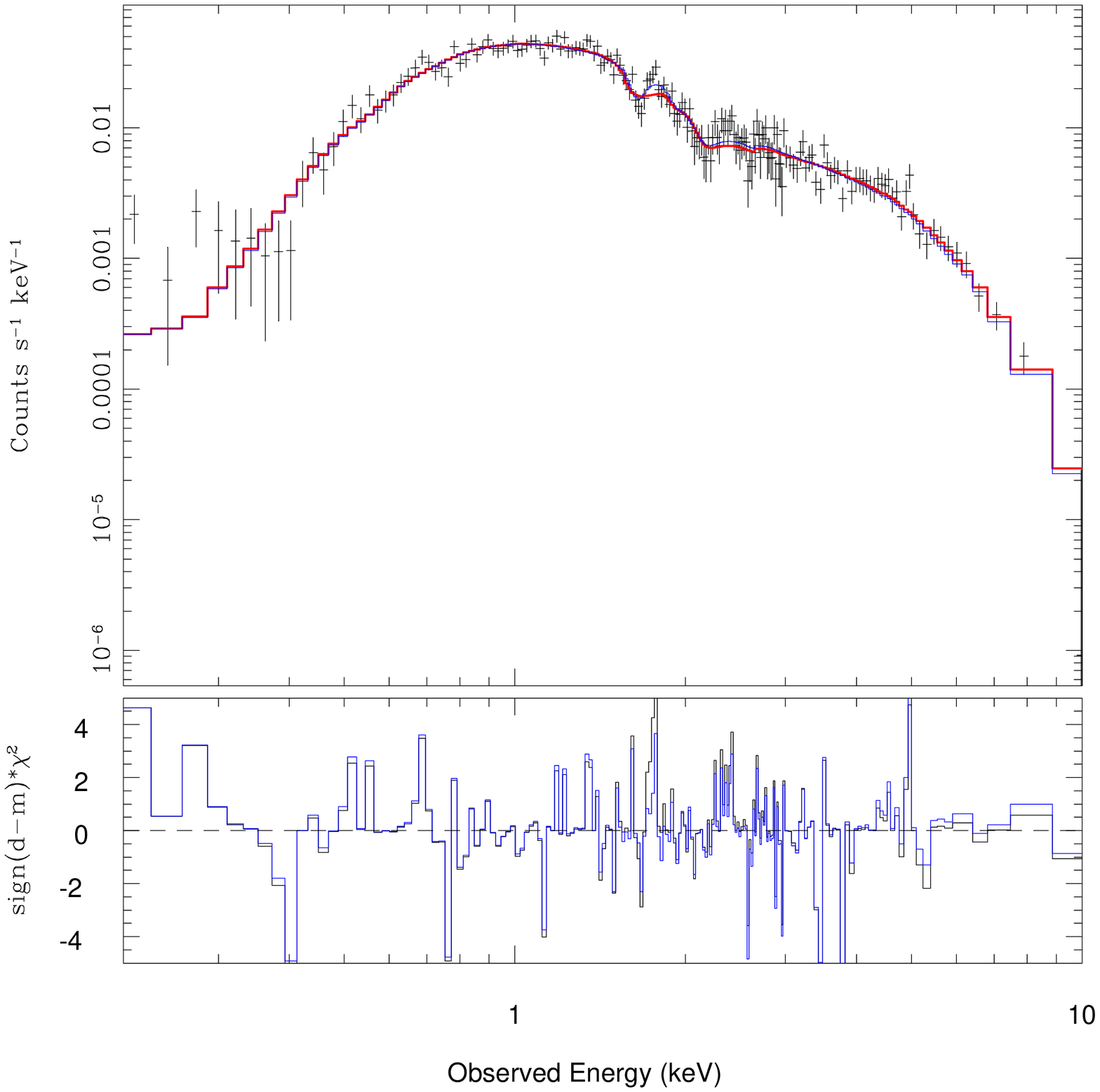}
\includegraphics{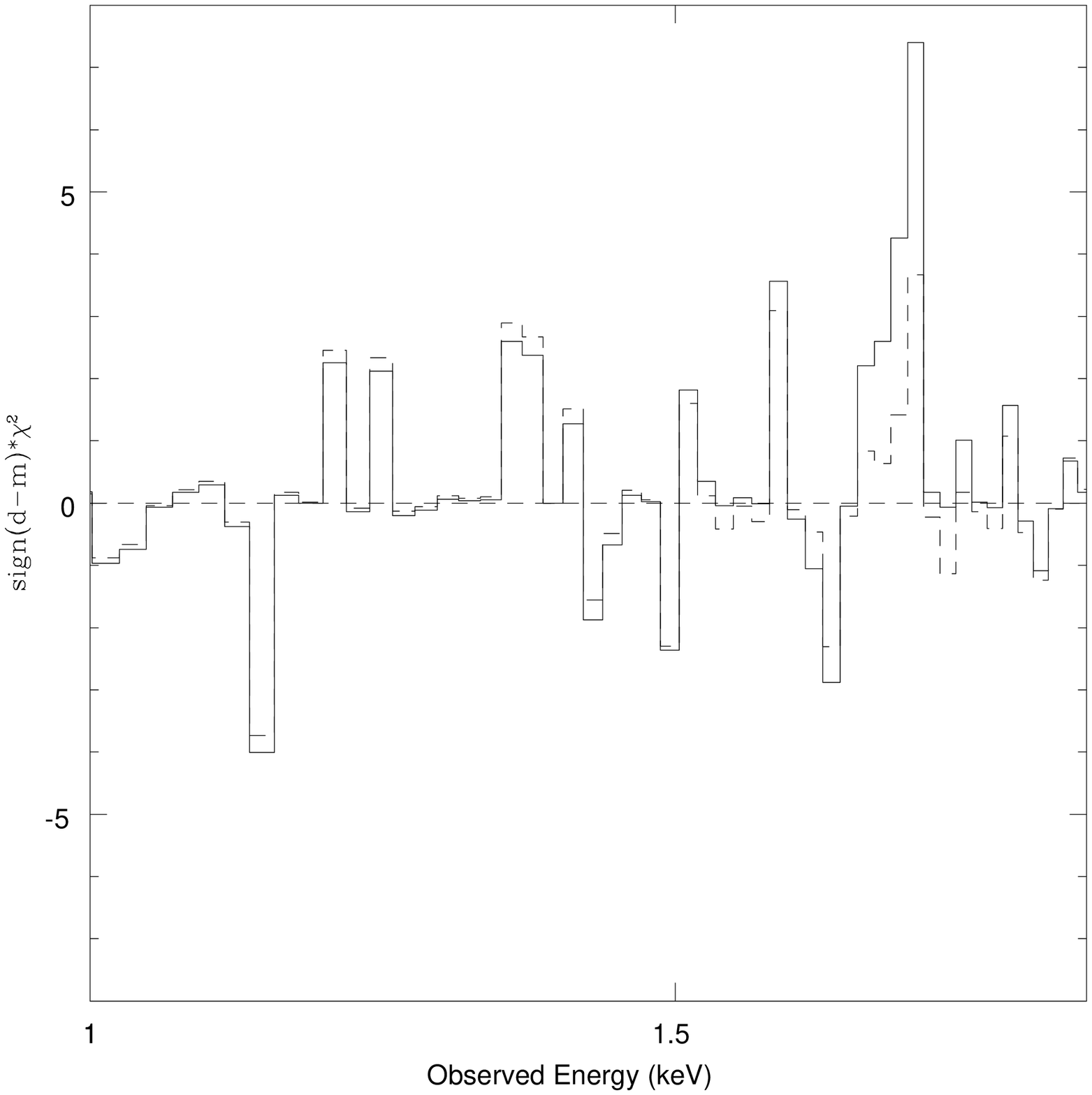}}
\caption{
\chan \sp X-ray spectrum of \apm \sp with two models over-plotted: the {\tt Edge} model (solid red line)
and the {\tt 2-Absorption-lines} model (solid blue line). In the right panel,
we show a closer view of the residuals in the range $1-2$ keV. Solid line is the {\tt Edge} model. Dashed line is the {\tt 2-Absorption-lines} model.
\label{chan_comp1}}
\end{figure}

First, we applied the simplest strategy: We use $\chi^2-$statistics to describe 					%
the goodness-of-fit of the data to the model. In Fig. \ref{chan_comp1} (left panel), we plot the {\tt Edge} model, 	%
taking the intrinsic absorption $N_H$, optical depth $\tau$ and energy of the edge from 				%
Table 2 of H02, leaving only the normalization and the photon index $\Gamma$ 						%
free to vary (all these numbers are fully consistent with our own best-fit parameters, see Table \ref{tbl1}). 		%
The resulting $\chi^2$ is 192 for 185 d.o.f. We also plot the {\tt 2-Absorption-lines} model, 				%
but this time we allow all parameters free to vary (all the numbers fully consistent with the parameters		%
reported by C02).													%
This time $\chi^2=179$ for 180 d.o.f. At this stage, we are unable to judge which model is better at describing 	%
the data, since it is true that the {\tt 2-Absorption-lines} model reports a smaller $\chi^2$, but it also 		%
introduces more free parameters in the fitting procedure, which might produce the advantage. A simple			%
hypotheses $F$-test is not permitted under this context, partly because these two {\it are not nested models}		%
\citep[a detailed discussion about conditions under which the $F$-test is valid is given in][]{protassov2002a}.		%
Therefore, because the differences between models are actually local, we focused on the spectral discrepancies close			%
to the strongest absorption feature at $\approx 1.6$ keV (observed frame), specifically the range $1-2$ keV.				%
We have adopted both models using the global best-fit parameters, but evaluate $\chi^2$ locally from 1 to 2 keV.			%
The results are: $\chi^2[{\rm Edge}]_{(1-2~{\rm keV})}=55$ and $\chi^2[{\rm 2Lines}]_{(1-2~{\rm keV})}=45$ for 52 d.o.f. First, it is	%
easy to see that the differences between globals $\chi^2$ (mostly) come from this spectral region.					%
The largest discrepancies between models are concentrated in the region around to 1.6 keV. We show these differences						%
in Fig. \ref{chan_comp1} (right panel). Nevertheless, we are still not able to answer the question: Which model (between these two) is the best to describe	%
the data? 																			%
                                                              %
                                                              %
                                                              %
                                                              %
%
This locates us in the context of hypotheses testing problems, and we will 	%
make use of the standard Bayesian solution to this problem computing        	%
the Bayes factor for one hypotheses against the other.                          %
The Bayes factor $B_{21}$ for a model $M_2$ against another   		%
model $M_1$ given the data $y$ is the ratio of posterior probability,	%
namely                                            			%
\begin{equation}                                              %
B_{21}=\frac{P(y|M_2)}{P(y|M_1)},                             %
\label{b21}
\end{equation}                                                %
the ratio of marginal likelihoods. We consider the estimation %
of the integrated likelihood from posterior simulations       %
output \citep{raftery2007a}.                                  %
This strategy is becoming popular for hypotheses testing in   		  		%
astrophysical contexts \citep[e.g.,][]{protassov2002a,gregory2005a,trotta2008a}.	%
Thus, we followed \cite{raftery2007a} and use the harmonic mean 		%
identity, which says that the reciprocal of the integrated    			%
likelihoods is equal to the posterior harmonic mean of the     			%
likelihoods.						      			%
The simplest estimator of the harmonic mean is:               %
                                                              %
                                                              %
\begin{equation}                                              				%
\pi_{HM}(y)=\left[ \frac{1}{B}\sum_{t=1}^B{\frac{1}{\pi(y|\theta^t)}}\right ]^{-1}	%
\end{equation}                                                				%
based on B draws $\theta^1,\theta^2,...\theta^B$ from the     				%
posterior distribution $\pi(\theta|y)$. This sample might come   			%
out of a standard Markov chain Monte Carlo implementation,    				%
for example.                                                  				%
Indeed, we make use of Monte Carlo Markov chain (MCMC) simulations 			%
to compute the marginal likelihoods necessary to confront one model against   		%
the other.                                                                      	%

{\it Simulation 1: 2-Absorption-lines and Edge models without    					%
constraint in their parameters $.-$} This means that, for each model, the MCMC routine is 		%
allowed to explore the full parameter space.								%
We run a MCMC with $B=2000$ on the \chan \sp data, using      				%
two models:                                                   				%
										
{\bf Model 1 .-} Power-law with intrinsic absorption and an absorption Fe Edge at 	%
$\approx 7.7$ keV (quasar frame).                                               	%

{\bf Model 2 .-} Power-law with intrinsic absorption and two absorption lines at 	%
$\approx 8$ keV and $\approx 9.5$ keV (quasar frame).                            	%

Finally, the specification of a statistical model requires    %
the form of the likelihoods terms. In our model, each         %
likelihood term is                                            %
\begin{equation}                                              					%
\frac{1}{\sqrt{2\pi}\sigma_i}\exp{\left [ -\frac{(M_i^k-y_i)^2}{2\sigma_i^2} \right ]},		%
\end{equation}                                                					%
where $y_i$ and $\sigma_i$ are the data (counts) and its      					%
uncertainty in the bin $i$.				      					%
In this context, $k=1,2$ refer evaluation of models 1 and 2.					%
The index $i$ runs over the bins falling into the range $1-2$ keV 				%
(52 bins for the \chan \sp data).				  				%

This concludes the statistical specification of our           %
simulation.                                                   %
                                                              %
A major drawback of the harmonic mean estimator is its        %
computational instability \citep{raftery2007a}. In fact,      %
by monitoring the cumulative harmonic mean of simulation 1,   %
we could see very large jumps, evidencing this instability \citep{raftery2007a}. %
In order to avoid statistical complications and make some     %
progress, we have constrained the energies of the two lines   %
in Model 2 and re-run a second simulation.                    %

{\it Simulation 2: 2-Absorption-lines and Edge models with    %
constraints in the line energies $.-$}			      %
This means that we run our MCMC with the energies of the two  %
lines in Model 2 fixed at 8 keV and 9.7 keV, values           %
taken from best-fit parameters of C02        		      %
and fully compatible with our own best-fit values.	      %
Here, again $B=2000$ and the models are the same as before.   %
                                                              %
                                                              %
                                                              %

The Monte Carlo simulations are run using parameter           %
values fitted to the data under the respective models         %
and account for uncertainty in these fitted values.           %
From the resulting $\chi^2(1-2~{\rm keV})$, we find that Model 2     	%
always fits the data better than Model 1.  		      		%
We check the stability of the harmonic means through        		%
monitoring. We note that the harmonic means for this 	      		%
simulation are stable.					      		%
Finally, we compute the Bayes factor of Model 2 against Model 1, 	%
using Eq. (\ref{b21}) and found $B_{21}\approx 22$ dB.			%
\footnote{Decibans (tenths of a power of 10), is a common unit to	%
represent weights of evidence.}						%
According to \cite{jeffreys1961a}, this can be interpreted as 		%
``{\it strong evidence}" for Model 2 against Model 1, given this data.	%
In our context, the {\tt 2-Absorption-lines} model better describes   	%
the $1-2$ keV spectrum of \chan, against      				%
the {\tt Edge} model.                                               	%

Now, we proceed to assess the goodness-of-fit, of the         %
two models presented, to the \xmm \sp data.            	      %
                                                              %
We applied exactly the same methodology previously described;  %
simulations 1 and 2 have the same meaning, and models 1 and 2, %
too. But the underlying data is the \xmm \sp data.             %
The results given by simulation 2   				 		%
(this time 35 bins are included) again show Model 2 producing  			%
smaller $\chi^2(1-2~{\rm keV})$,  					 	%
although 1) the difference is much smaller compared to Model 1 			%
[only $\Delta \chi^2(1-2~{\rm keV})\approx 3$]; and 2) the parameters are 	%
less constrained (compared with the same simulation using \chan	\sp data).	%
The computation of the Bayes factor this time 			%
gives $B_{21}\approx 5$ dB. In the Jeffreys' scale, this is   	%
interpreted as {\it``barely worth mentioning"}.		      	%
This means that the \xmm \sp data is not able to discriminate	%
between the two models.						%
From here, it is clear that neither of the two models ({\tt Edge}	%
or {\tt 2-Absorption-lines} model) can be applied, with equal 	%
success (and unambiguously) to both sets of data.		%
In the next section, we will try to establish the physical 	%
scenario under such spectral modeling is possible, 		%
having as goal to propose one unified model that can reasonably %
present good  fits on both  spectra.				%

\section{Photoionization modeling of the X-ray Spectrum of \apm
\label{photo}}

We performed a photoionization modeling of the X-ray spectrum
of \apm, using the code XSTAR
\footnote{Version 2.1kn6. \\
See http://heasarc.gsfc.nasa.gov/docs/software/xstar/xstar.html.}
with the atomic data
of \cite{bautista2001a}. The code includes all the relevant atomic processes
(including inner shell processes) and computes the emissivities and optical depths
of the most prominent X-ray and UV lines identified in AGN spectra.
For that purpose,
we built a grid of photoionization models with the column density
of the ionized material ($N_H$), ionization parameter ($\log[\xi]$),
and Fe abundance 
\footnote{Atomic abundances are entered {\it relative to solar abundances} as defined in
\cite{grevesse1996a}, with $1.0$ being defined as the \emph{solar}.}
as variables.
Our models are based on spherical shells illuminated
by a point-like X-ray continuum
source.
The input parameters are the
source spectrum, the gas
composition, the gas density
$n_H$, the column density and
the ionization parameter.
The source spectrum
is described by the spectral luminosity
$L_{\epsilon}=L_{\rm ion}f_{\epsilon}$,
where $L_{\rm ion}$ is the integrated
luminosity from
1 to 1000 Ryd,
and $\int_{1}^{\rm 1000~Ryd} f_{\epsilon}
d\epsilon=1$. The spectral function
is taken to be a power-law
$f_{\epsilon}\sim \epsilon^{-\alpha}$,
and $\alpha$ is the energy index, equal to 1.
The spectra contain edges and absorption
lines from
the following
elements, H, He,
C, N, O, Ne, Mg, Si, S,
Ar, Ca, and Fe. We use
the abundances of \cite{grevesse1996a}
in all our models (we use the term {\it solar} for these abundances).
We adopt a turbulent velocity
of 1000 \kms, and a hydrogen density of $10^{12}$ \cmd, as input parameters.

\subsection{Single-Absorber photoionization model}

Our first step was to see how well a {\tt single-absorber}
photoionization model can reproduce
the broad-band X-ray spectrum of \apm, without any prior assumption about
any of the parameters of interest, $N_H$, $\log(\xi)$, outflow velocity
($v_{out}$), or abundances. So, we started by fitting a solar abundance model at
rest in the frame of the quasar ($v_{out}=0$ \kms). We call this Model A.
We apply the same photoionization model to both sets of data, the \xmm \sp
and \chan \sp X-ray observation of \apm. The best-fit column density is
$(1.14 _{-0.06} ^{+0.07})\times 10^{23}$ \cmn \sp
($[1.32 _{-0.10} ^{+0.14}]\times 10^{23}$ \cmn) 
with $\log(\xi)=1.50 _{-0.05} ^{+0.06}$ ($\log[\xi]=1.50 _{-0.07} ^{+0.08}$),
for the \xmm \space (\chan) data. The high column density is in agreement with
previous fitted column densities to this spectrum \citep[e.g.,][]{hasinger2002a},
and with the general trend of high column densities observed in high-redshifted
quasars \cite[e.g.,][]{gallagher2002a}. The goodness-of-fit is measured with
$\chi^2$, equal to 339.7 for 289
d.o.f for the \xmm \space data, and  $\chi^2 =241.3$ (with 183 d.o.f)
for the \chan \sp data.
These global fits are statistically unacceptable             %
\footnote{$P_{289,339.7}[{\rm Model~A};\xmm]=2 \times 10^{-2}$ and  %
$P_{183,241.3}[{\rm Model~A};\chan]=2 \times 10^{-3}$, where        %
$P_{\nu,\chi^2}$ is the probability of exceeding $\chi^2$ for %
$\nu$ degrees of freedom.}. 				      %
Now, as we showed in the previous section, we need some local 	%
criterion to assess the quality of the fits in the region   	%
$1-2$ keV. Therefore, we have adopted the global parameters of  %
Model A and compute $\chi^2$ (locally) from 1 to 2 keV.   %
The results are: $\chi^2_{(1-2~{\rm keV})}[\xmm]=79$ for 35 d.o.f,  %
and $\chi^2_{(1-2~{\rm keV})}[\chan]=91$ for 52 d.o.f. These fits   %
are rejected even locally (see Table \ref{tbl2} for global    %
and local $\chi^2$ and $P_{\nu,\chi^2}$).                     %
                                                              %
                                                              %
                                                              %
                                                              %
                                                              %
                                                              %

Motivated by the interpretation of H02 that the absorption
feature around 7.7 keV (rest-frame) may be a Fe edge formed by bound-free
transitions of iron ions from Fe~{\sc xv} to Fe~{\sc xviii}, and by the fact that
the fitted edge would require Fe/O $\sim 2-5$, we have fitted our model to both sets of data
leaving the Fe abundance free to vary. Again, we use a rest-frame approach (Model B).
The result is shown in Table \ref{tbl2}. We can see that this model provides
a good description of the {\it global} spectrum, $\chi^2 =310.1$ (for 288 d.o.f) for
the \xmm \sp fit and $\chi^2 =204.1$ (for 182 d.o.f) for the \chan \sp fit.
We note a relatively important decrease of the column density of the ionized
material, now $N_H\sim 5-6 \times 10^{22}$ \cmn. This is owing (in part) to the fact that
the spectrum has strong spectral features produced by iron (in the hard X-ray band); 
for example, the edge composed by ionized species of Fe from Fe~{\sc xv} to Fe~{\sc xviii},
and likely Fe spectral lines. The ionization parameter remained almost the
same.

\begin{figure}
\rotatebox{-90}{\resizebox{7cm}{!}{
\includegraphics{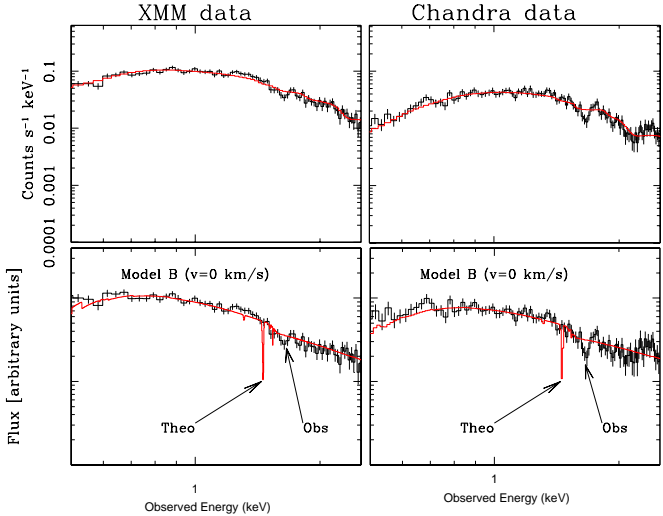}}}
\caption{
Instrument convolved (top), and physical model (bottom) of Model B applied to the \xmm \sp
data (left) and \chan \sp data (right).
\label{modelb1}}
\end{figure}

As we see in Table \ref{tbl2}, Model B global-fits are fairly %
good (i.e, acceptable $P_{\nu,\chi^2}$) but a closer look     %
at the region ($1-2$) keV reveals important discrepancies     %
between the model and the data.				      %
This difference can easily be seen in Fig. \ref{modelb1}.     				%
In the top panels, we have the {\it instrument convolved} best-fit Model B ($v=0$ \kms)	%
to each of the data sets, and in the bottom panels we present 				%
the flux spectra with the model without convolution with the  				%
instrument response, so we can see physically where the strongest 			%
absorption features are predicted to be.		      				%
In fact, there is an important mismatch between theory and     				%
data. The statistical evidences can easily be read out from    				%
Table \ref{tbl2}. The local $\chi^2_{(1-2~{\rm keV})}$  for \xmm    			%
\sp (\chan) is 55.5 for 35 d.o.f (65.5 for 52 d.o.f), resulting				%
in $P_{(1-2)~{\rm keV}}[{\rm Model~B};\xmm]=0.01$ and		      			%
$P_{(1-2)~{\rm keV}}[{\rm Model~B};\chan]=0.1$. These probabilities locate 		%
Model B, very close to the rejection limit  						%
(a common threshold to reject the null hypotheses is          				%
$P_{\nu,\chi^2}\le0.05$).				      				%
\begin{table}
\caption{XSTAR fit results and comparison between sets of data.
{\it XMM-Newton vs Chandra} observation of \apm.}
\label{tbl2}
\centering
\begin{tabular}{l c c}
\hline \hline
Parameter&
XMM data&
Chandra data
\\
\hline
\multicolumn{3}{c}{\small Model A: Fe fixed to solar; 0 \kms}\\ \hline

$\Gamma$     		&   $2.11 _{-0.03} ^{+0.03}$                  &  $1.84 _{-0.03} ^{+0.03}$  \\
Norm$^{\rm a}$         	&   $1.34 _{-0.02} ^{+0.02} \times 10^{-4}$   &  $1.08 _{-0.02} ^{+0.03} \times 10^{-4}$  \\
$N_H$ (\cmn) 		&   $1.14 _{-0.06} ^{+0.07} \times 10^{23}$   &  $1.32 _{-0.10} ^{+0.14} \times 10^{23}$  \\
$\log(\xi)$  		&   $1.50 _{-0.05} ^{+0.06}$                  &  $1.50 _{-0.07} ^{+0.08}$                 \\
$\chi^2$/(d.o.f)    				& 339.7/(289)                     &  241.3/(183)                       \\
$P_{{\rm global}}[{\rm Model~A}]$		& $2 \times 10^{-2}$		  & $2 \times 10^{-3}$		       \\
$\chi^2(1-2~{\rm keV})$/(d.o.f)       		& 79/(35)                     	  &  91/(52)         	               \\
$P_{(1-2~{\rm keV})}[{\rm Model~A}]$        	& $4 \times 10^{-6}$              & $3 \times 10^{-4}$                 \\

\hline
\multicolumn{3}{c}{\small Model B: Fe free to vary; 0 \kms}\\ \hline

$\Gamma$            	& $2.06 _{-0.03} ^{+0.03}$                  & $1.82 _{-0.03} ^{+0.03}$ 			\\
Norm$^{\rm a}$      	& $1.32 _{-0.02} ^{+0.03} \times 10^{-4}$   & $1.11 _{-0.02} ^{+0.02} \times 10^{-4}$  	\\
$N_H$ (\cmn)        	& $5.08 _{-0.13} ^{+0.20} \times 10^{22}$   & $5.97 _{-0.25} ^{+0.38} \times 10^{22}$  	\\
$\log(\xi)$         	& $1.50 _{-0.18} ^{+0.03}$                  & $1.50 _{-0.13} ^{+0.05}$  			\\
Fe        		& $5.04 _{-0.23} ^{+0.44}$                  & $4.83 _{-0.27} ^{+0.63}$  			\\
$\chi^2$/(d.o.f)  				& 310.1/(288)                   & 204.1/(182) 					\\
$P_{{\rm global}}[{\rm Model~B}]$           	& $0.19$              		& 0.14                 				\\
$\chi^2(1-2~{\rm keV})$/(d.o.f)       		& 55.5/(35)                     &  65.5/(52)                           		\\
$P_{(1-2~{\rm keV})}[{\rm Model~B}]$        	& 0.01              		& 0.10                 				\\

\hline
\multicolumn{3}{c}{\small Model C: Fe fixed to solar; outflow at 0.21c}\\ \hline

$\Gamma$     		&   $2.12 _{-0.03} ^{+0.03}$                  &  $1.83 _{-0.03} ^{+0.03}$  \\
Norm$^{\rm a}$         	&   $1.35 _{-0.02} ^{+0.02} \times 10^{-4}$   &  $1.06 _{-0.02} ^{+0.02} \times 10^{-4}$  \\
$N_H$ (\cmn) 		&   $7.08 _{-0.27} ^{+0.30} \times 10^{22}$   &  $6.72 _{-0.36} ^{+0.41} \times 10^{22}$  \\
$\log(\xi)$  		&   $1.82 _{-0.04} ^{+0.04}$                  &  $1.72 _{-0.09} ^{+0.08}$                 \\
$\chi^2$/(d.o.f)    				& 351.2/(289)                     &  246.5/(183)                       \\
$P_{{\rm global}}[{\rm Model~C}]$           	& $7 \times 10^{-3}$              & $1 \times 10^{-3}$                 \\
$\chi^2(1-2~{\rm keV})$/(d.o.f)       		& 89/(35)                     	  &  101.2/(52)                        \\
$P_{(1-2~{\rm keV})}[{\rm Model~C}]$      	& $5 \times 10^{-8}$          	  & $1 \times 10^{-5}$                 \\

\hline
\multicolumn{3}{c}{\small Model D: Fe free to vary; outflow at 0.21c}\\ \hline

$\Gamma$            & $2.08 _{-0.03} ^{+0.03}$                  & $1.85 _{-0.03} ^{+0.03}$ 			\\
Norm$^{\rm a}$      & $1.32 _{-0.02} ^{+0.02} \times 10^{-4}$   & $1.13 _{-0.03} ^{+0.02} \times 10^{-4}$  	\\
$N_H$ (\cmn)        & $3.85 _{-0.08} ^{+0.08} \times 10^{22}$   & $3.71 _{-0.08} ^{+0.09} \times 10^{22}$  	\\
$\log(\xi)$         & $1.78 _{-0.06} ^{+0.06}$                  & $2.12 _{-0.04} ^{+0.04}$  			\\
Fe        	    & $4.84 _{-0.28} ^{+0.30}$                  & $8.36 _{-0.49} ^{+0.53}$  			\\
$\chi^2$/(d.o.f)    				& 333.7/(288)           & 207/(182) 					\\
$P_{{\rm global}}[{\rm Model~D}]$           	& 0.03             	& 0.11                 				\\
$\chi^2(1-2~{\rm keV})$/(d.o.f)       		& 71.4/(35)             & 62.6/(52)                        		\\
$P_{(1-2~{\rm keV})}[{\rm Model~D}]$        	& $9 \times 10^{-5}$    & 0.17         		       			\\
\hline
\end{tabular}
\begin{flushleft}
The error parameters are 90\% confidence limits.
(a) Power-law normalization, photons keV$^{-1}$cm$^{-2}$s$^{-1}$ at 1 keV
in the observed frame.
\end{flushleft}
\end{table}
Based on the last result, we went further and explored the possibility that the gas absorbing X-rays
in \apm \sp is outflowing at intermediate-to-relativistic velocities. For that
purpose we have shifted the spectra 
\footnote{i.e., not taking into account radiative transfer effects.},
produced with our XSTAR-based ionization models,
by an array of velocities, from 0.08$c$ to 0.30$c$, with 0.01$c$ of resolution, and fit
in several ways the X-ray spectrum of \apm, using the \xmm \sp and \chan \sp data.
\footnote{We made the analysis of best-fit velocity by inspecting the evaluation of 
$\chi^2$ at every outflow velocity point of our grid of velocities, while the other parameters 
of interest are varied as usual. Then, we proceed to adopt the velocity-model
with the minimum $\chi^2$.}

\begin{figure}
\resizebox{\hsize}{!}{\includegraphics{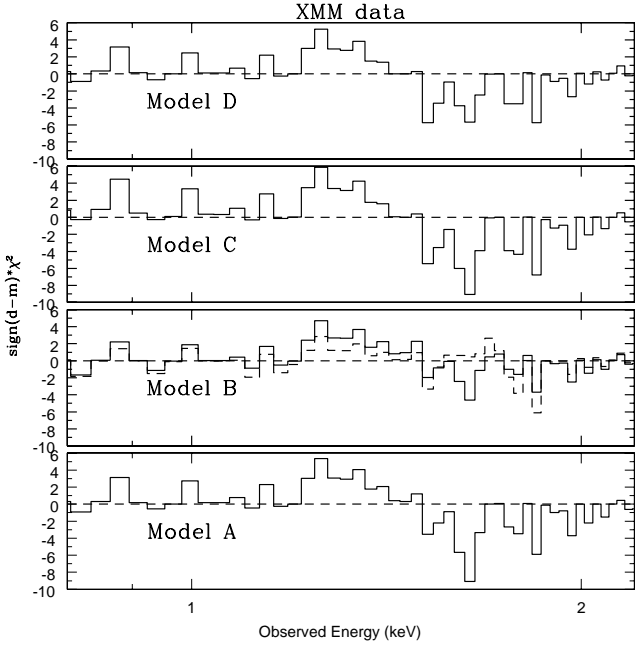}}
\caption{
Residuals (in sign(data-model)$\chi^2$) of our photoionization models A,B,C and D to the \xmm \sp data, in the band
$(1-2)$ keV observed frame. Dashed line is the residual of the {\tt two-absorbers} model.
See text.
\label{xmm_abcd}}
\end{figure}

\begin{figure}
\resizebox{\hsize}{!}{\includegraphics{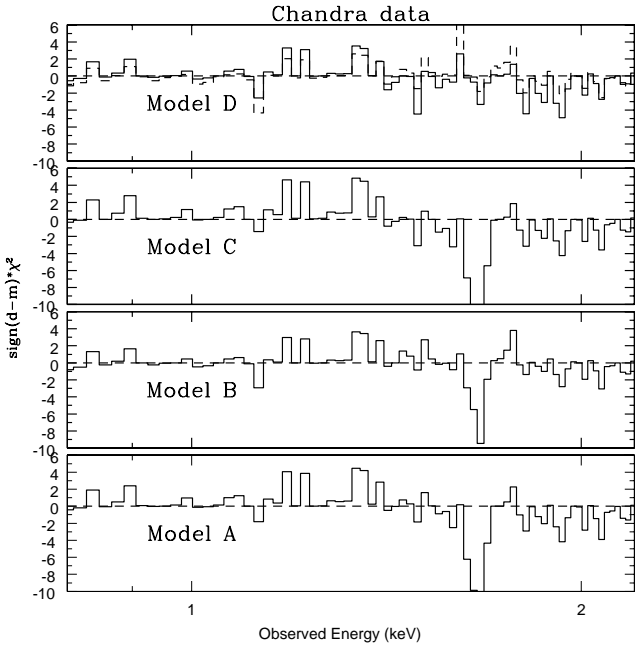}}
\caption{
Residuals (in sign(data-model)$\chi^2$) of our photoionization models A,B,C and D to the \chan \sp data, in the band
$(1-2)$ keV observed frame. Dashed line is the residual of the {\tt two-absorbers} model.
See text.
\label{chan_abcd}}
\end{figure}

Table \ref{tbl2}, quotes the best-fit parameters of the four most interesting fits,
in the context of the {\tt single-absorber} model. Models A and B
are models with Fe \emph{solar} and Fe \emph{free} to vary at $v_{out}= 0$ \kms, respectively. 
Then, we have
Model C and Model D, with Fe \emph{solar} and Fe \emph{free} (respectively) at
$v_{out}= 0.21c$, for both sets of data, \xmm \sp and \chan.
The {\tt single-absorber} model cannot be ruled out (instantaneously), but it does
not give a consistent fit to both sets of data (Model B for \xmm \sp and Model D for \chan).
Therefore, we explore the possibility of a multi-component photoionization model.
In Figs. \ref{xmm_abcd} and \ref{chan_abcd}, we plot the residuals between each model presented
here and the data. The dashed lines are residuals from a {\tt two-absorbers} model, which we
discuss in the next section. A detailed physical discussion
of these scenarios and its implications are given in Sect. \ref{discuss}.

\subsection{Two-Absorbers photoionization model}

Now that we have explored how the {\tt single-absorber} model  	%
fits the data, we can go further, and see if the data         	%
supports a multicomponent photoionization model.        	%
The simplest such model is a two-photoionized-absorbers model, 		%
and we investigate if the addition of an extra component      		%
to the best-fit {\tt single-absorber} model is statistically   		%
significant. This {\tt two-absorbers} model consists of:	      	%
one component at $v=0$ \kms \sp (rest-velocity component), 		%
and one at $v=0.18c$ (high-velocity component) 				%
\footnote{This is the best high-velocity component we    		%
found able to fit both sets of data with high $\chi^2-$probability.}. 	%
We take the best {\tt single-absorber} model (from Table \ref{tbl2}),  	%
selected as the best combination between global     		      	%
and local $P_{\nu,\chi^2}$. Model B is the best for \xmm \sp data     	%
and Model D is the best for the \chan \sp set. The addition of	      	%
an extra component (high-velocity component for \xmm \sp and          	%
rest-velocity component for \chan) significantly improves the fit     	%
compared to the {\tt single-absorber} model at the greater than        	%
99.9\% confidence level in both cases (according to the $F$-test),      %
both globally and locally.				      		%

\begin{figure}
\rotatebox{-90}{\resizebox{6cm}{!}{\includegraphics{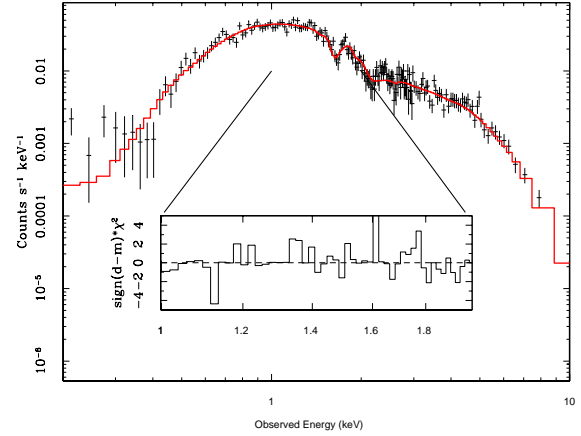}}}
\caption{
\chan \sp X-ray spectrum of \apm \sp in the $0.2-10$ keV (observed frame) band.The solid thick line is the best-fit {\tt two-absorbers} model.
The inside plot shows the residual (in sign(data-model)$ \chi^2 $) for the $1-2$ keV band. This model gives good agreement on both sets of data.
\label{xmm_two}}
\end{figure}

\begin{figure}
\rotatebox{-90}{\resizebox{6cm}{!}{\includegraphics{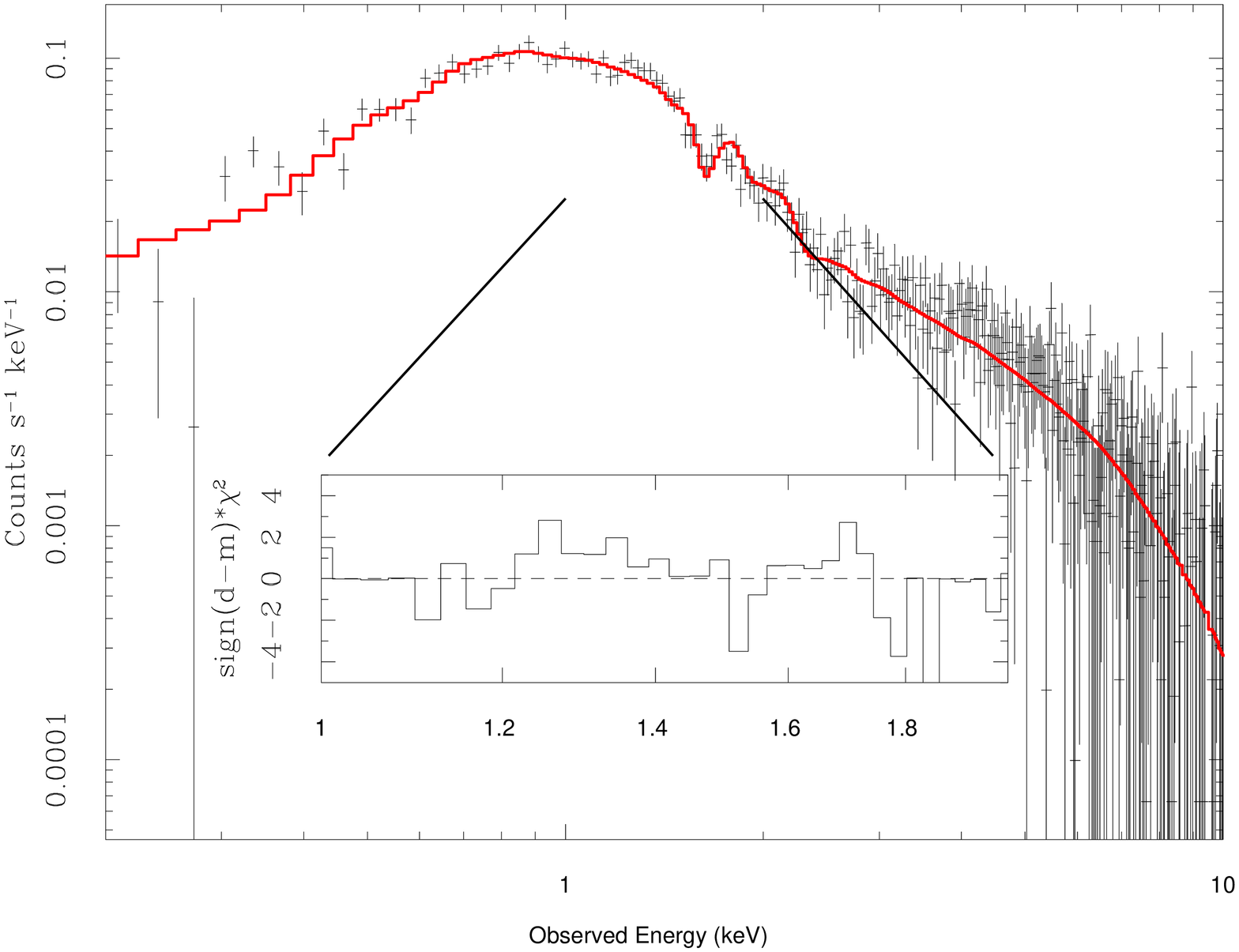}}}
\caption{
\xmm \sp X-ray spectrum of \apm \sp in the $0.2-10$ keV (observed frame) band.The solid thick line is the best-fit {\tt two-absorbers} model.
The inside plot shows the residual (in sign(data-model)$\chi^2$) for the $1-2$ keV band. This model gives good agreement on both sets of data.
\label{chan_two}}
\end{figure}

First, we fit a {\tt two-absorbers} model to each set of data 		%
separately and then we do it simultaneously, to    			%
check for inconsistencies between fits. 		      		%
In Table \ref{tbl4}, we have the results for these three       		%
schemes (Cols. 2,3,4). Figures \ref{xmm_two} and \ref{chan_two}		%
present plots of the best-fit {\tt two-absorbers} model over		%
the \chan \sp and \xmm \sp data, respectively.				%
The fits between data sets give      					%
different best-fit parameters at $\gtrsim 1\sigma$, for six out  	%
of the seven fitted parameters ($\log\xi[2]$ is fully consistent). 	%
The {\tt two-absorbers} model gives acceptable fits with      		%
high $\chi^2-$probability for both sets of data, but there     		%
exist small differences between best-fit parameters.          		%
\begin{table*}
\caption{Two-Absorber fits for \xmm \sp and \chan.}
\label{tbl4}
\centering
\begin{tabular}{r r r r r r r r}
\hline \hline
Parameter&   	
XMM&   		
Chandra&   	
Simult.&   	
XMM&   		
Chandra& 	
XMM&            
Chandra\\       
\hline

$\Gamma$     		&   $2.08 _{-0.03} ^{+0.03}$   &  $1.84 _{-0.02} ^{+0.02}$ & $2.00 _{-0.01} ^{+0.01}$& \ldots& \ldots & $2.13_{-0.02}^{+0.02}$ & $1.83 _{-0.02} ^{+0.02}$ \\
Norm$^{\rm a}$      	&   $1.41 _{-0.03} ^{+0.03}$   &  $1.18 _{-0.02} ^{+0.02}$ & $1.34 _{-0.01} ^{+0.01}$& \ldots& \ldots & $1.50_{-0.02}^{+0.02}$ & $1.17 _{-0.02} ^{+0.02}$\\
$N_H(1)$$^{\rm b}$     	&   $5.72 _{-0.19} ^{+0.21}$   &  $6.05 _{-0.19} ^{+0.19}$ & $7.34 _{-0.15} ^{+0.18}$& \ldots& \ldots & \ldots & \ldots\\
$\log\xi(1)$ 		&   $1.15 _{-0.15} ^{+0.14}$   &  $1.50 _{-0.12} ^{+0.03}$ & $1.50 _{-0.04} ^{+0.02}$& \ldots& \ldots & \ldots & \ldots\\
Fe        		&   $3.14 _{-0.25} ^{+0.27}$   &  $4.79 _{-0.22} ^{+0.23}$ & $2.85 _{-0.10} ^{+0.10}$& \ldots& \ldots & \ldots & \ldots\\
$N_H(2)$$^{\rm b}$     	&   $4.75 _{-0.89} ^{+1.20}$   &  $2.74 _{-1.15} ^{+0.68}$ & $5.49 _{-0.54} ^{+0.71}$& \ldots& \ldots & \ldots & \ldots\\
$\log\xi(2)$         	&   $3.03 _{-0.04} ^{+0.15}$   &  $3.10 _{-0.12} ^{+0.23}$ & $3.00 _{-0.01} ^{+0.04}$& \ldots& \ldots & \ldots & \ldots\\
$\chi^2$/(d.o.f)    	&   296.1/(286)                &  186.5/(180)              & 647.9/(473) 	     & 373.8/(293) 	    & 274.1/(187)       & 298.5/(291) & 189.3/(185) \\
$P_{\rm global}[{\rm Two~Abs}]$		&   0.35	       &  0.38		       	   & $4\times 10^{-8}$	     & $7\times 10^{-4}$ & $2\times 10^{-5}$ & 0.39	      & 0.43 \\
$\chi^2(1-2~{\rm keV})$/(d.o.f)       	&   44.9/(35)          &  48.9/(52)   	           & 135.1/(87)		     & 65.7/(35) 	 & 69.4/(52)         & 43.9/(35)      & 49.3/(52) \\
$P_{(1-2~{\rm keV})}[{\rm Two~Abs}]$   	&   0.14               &  0.65  		   & $4\times 10^{-4}$ 	     & $7\times 10^{-4}$ & 0.05 	     & 0.17	      & 0.63 \\
\hline
\end{tabular}
\begin{flushleft}
The error parameters are 90\% confidence limits.
The Fe abundance is the same for both absorbers.
(a) Power-law normalization, $\times 10^{-4}$ photons keV$^{-1}$cm$^{-2}$s$^{-1}$ at 1 keV
in the observed frame.
(b) Column density of the component, $\times 10^{22}$ \cmn.
\end{flushleft}
\end{table*}

If we fit both data sets simultaneously (Col. 4), we find  		%
reasonable consistency between them and the separated fits.  		%
The most notable discrepancy is seen in the power-law			%
component, with differences of $\sim 10-15$\% in its parameter values. 	%
To check for this difference, we have taken all the parameters 		%
resulting from the simultaneously fit and fixed them to      		%
each set of data separately (Cols. 5,6), and compute        		%
$\chi^2-$probabilities. In both cases, the fits are rejected 		%
(in both global and local senses).			      		%
Finally, we have taken the simultaneous best-fit parameters  		%
and fixed them, except that we allowed the power-law          		%
parameters free to vary in each set of data (Cols. 7,8).    		%
We recover the goodness-of-fit and the model becomes acceptable. 	%
Then, we compute integrated observed fluxes
(later in Sect. \ref{discuss} we also compute intrinsic luminosities)
using both set of data in the band $0.2-10$ keV
with the following results: $F_{(0.2-10~{\rm keV})}[\chan]=6.9\pm 0.3
\times 10^{-13}$ \flux and $F_{(0.2-10~{\rm keV})}[\xmm]=7.6\pm 0.3
\times 10^{-13}$ \flux, thus the differences seen in the power-law
component are reflected in a change of $\sim 10$ \% on fluxes,
\footnote{For completeness we have computed the observed flux of the
$\sim 16$ ks \xmm \sp observation (XMM1) and the result is:
$F_{(0.2-10~{\rm keV})}[{\rm XMM1}]=9.6\pm 0.6 \times 10^{-13}$ \flux.}
and this could produce
small changes in the physical parameters of the absorbers
(the most notorious are $\log\xi(1)$, $\approx$ 30\% and
$N_H(2)$, $\approx$ 70\%). However, the errors on the parameters allow us
to obtain high  $\chi^2-$probability if, appart from the power-law,
both observations are fitted with the same physical two absorbers,
opening the possibility that {\it both absorbers have been there in both observations}.

\section{Line Identification}

We are now in position of studying in more detail the possible			%
identification of the feature $\sim 8$ keV (rest-frame) of the X-ray		%
spectrum of \apm. 								%
We will discuss two cases: i) the possible identification     			%
if we assume the {\tt single-absorber} model is the best, and ii) the possible 	%
identification if we consider the {\tt two-absorbers} model.       		%


\subsection{Case (i)}

Given the good agreement between the \chan \sp data
and our {\tt single-absorber} model at $v\sim 0.21c$ 
(Model D of Table \ref{tbl2}), we want to  investigate how well 
constrained is the best-fit ionization parameter found and, 
its ability to reproduce the feature $\sim 8$ keV as well as the X-ray 
absorption at lower energies.
For that purpose, we have taken one model in which the
column density $N_H$ is fixed to $5\times 10^{22}$ \cmn, built a grid
in $\xi$, $1.5\le \log(\xi) \le 4.5$ with resolution of 0.1, leaving the
iron abundance free to vary, and computed $\chi^2$ produced by fitting
our models at $v=0.21c$. The main result of this experiment is that clearly
the \chan \sp data strongly favors $\log(\xi)\approx 2.1 \pm 0.1$. 
For instance, at $\log(\xi) \approx 2.4$, there exists a clear 
deviation of the best-fit model from the data with 
$\Delta \chi^2 \approx 70$, translated in that the model becomes
rejected (from accepted). 
Thus, in Model D, the ionization state is mostly driven by the absorption at 
low energies, with an important contribution of the feature 
$\sim 8 $ keV. Under these physical conditions, the feature is 
better represented by a complex of lines produced by L transitions of
highly-ionized species of iron from Fe~{\sc xviii} to Fe~{\sc xxii}
(Fe C2 in Table \ref{tbl3}).
The centroid of this complex is located $\sim $ 8.1 keV,
i.e., $v\approx 0.21(\pm0.01)c$.
\begin{table*}
\caption{XSTAR spectral lines prediction.}
\label{tbl3}
\centering
\begin{tabular}{c c c c c c c}
\hline \hline
Complex&
Ion&
Transition&
$\lambda_{\rm lab}$&
$\tau_0(\log[\xi]\approx 3)$&
$\tau_0(\log[\xi]\approx 2.8)$&
$\tau_0(\log[\xi]\approx 2.1)$
\\
\hline
Fe C1 at            		&  Fe~{\sc xxv}    & 1$s^2~{\rm ^1S}-1s2p~{\rm ^1P}$ 	  	& 1.85 & 110	& 44	  	& $\ll 10^{-2}$ \\
$N_H=5\times 10^{22}$ \cmn      &  Fe~{\sc xxiv}   & 1$s^22s~{\rm ^2S}-1s2s2p~{\rm ^2P}$   	& 1.86 & 33	& 36     	& $\ll 10^{-2}$ \\
				&  Fe~{\sc xxiii}  & 2$s^2~{\rm ^1S}-1s2p~{\rm ^1P}$   		& 1.87 & 45	& 108     	& $\ll 10^{-2}$ \\

\hline
\hline
Fe C2 at                        &  Fe~{\sc xxii}   & 2$s^22p-1s2s^22p^2~{\rm ^2P}$              & 1.88 &  & 42             & 3  \\
$N_H=5\times 10^{22}$ \cmn      &  Fe~{\sc xxi}    & 2$p^2-1s2s^22p^3~{\rm ^3S}$                & 1.89 &  & 10             & 5  \\
                                &  Fe~{\sc xx}     & 2$s^22p^3-1s2s^22p^4~{\rm ^4P}$            & 1.91 &  & 3              & 18 \\
                                &  Fe~{\sc xix}    & 2$s^22p^4-1s2s^22p^5~{\rm ^3P}$            & 1.92 &  & 0.4            & 26 \\
                                &  Fe~{\sc xviii}  & 2$s^22p^5-1s2s^22p^6~{\rm ^2S}$            & 1.93 &  & 0.1            & 34 \\
\hline
\end{tabular}
\begin{flushleft}
Two different complex of lines are predicted at different ionization states.
The laboratory wavelength $\lambda_{\rm lab}$ is given in \AA.
\end{flushleft}
\end{table*}

\begin{figure}
\resizebox{\hsize}{!}{\includegraphics{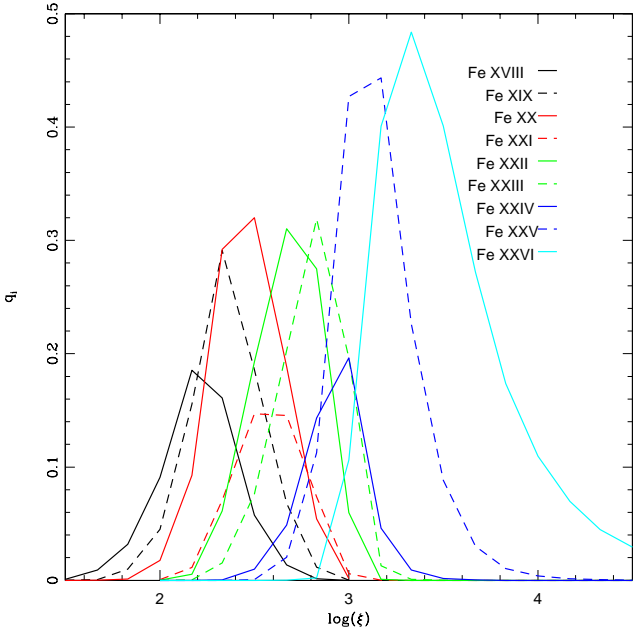}}
\caption{
XSTAR ionization balance computation (ionization fraction $q_i$).
Only shown in this plot species of Fe~({\sc xviii-xxvi}). The {\tt single-absorber} model favors
$\log(\xi)\sim 2.1$, and the absorption feature at $\sim$8 keV rest-frame would be
identified as the complex Fe C2 in Table \ref{tbl3}. However, the HV component of the {\tt two-absorbers} model
strongly suggest that the feature can be associated to the \ion{Fe}{xxv} K$_\alpha$ line
($\log(\xi)\sim 3$ is precisely where Fe$^{+24}$ has its peak).
\label{fe_fracc}}
\end{figure}

This identification is different from the one made by C02,
who identify this feature as Fe~{\sc xxv} K lines. That identification
{\it is not based on a photoionization model}, but on searching for the closest
(and most conservative) line (with some other physical arguments), in
likely several atomic data bases. The problem with the identification of this
feature as Fe~{\sc xxv} lines (in our Model D), is that they required that
the plasma be at $\log(\xi)\gtrsim 2.8$. 
Only at this highly-ionized state, the ionization fraction of ions 
like Fe$^{+22}$, Fe$^{+23}$ and Fe$^{+24}$, are high enough to form spectral lines. 
In that physical scenario, a complex of lines (Fe C1), produced by transitions 
of iron ions from Fe~{\sc xxiii} to Fe~{\sc xxv}, arises naturally, 
where the K lines of Fe~{\sc xxv} are involved. 
However, our Model D at this $\log(\xi)\approx 2.8$ is a bad 
representation of the absorption at lower energies. 
In Fig. \ref{fe_fracc}, we present the ionization fraction ($q_i$) 
of Fe ions from Fe~{\sc xviii} to Fe~{\sc xxvi} of our grid of XSTAR
photoionized clouds at different $\log(\xi)$. At $\log(\xi)\approx 2.1$
($\sim$ best-fit ionization parameter) the predominant ions are Fe ions
from Fe~{\sc xviii} to Fe~{\sc xxii}, and the ionization fraction of
Fe$^{+24}$ ($q_{\rm FeXXV}), is\sim 10^{-7}$. 
In Table \ref{tbl3} we present:
the transitions contributing to the complex, the laboratory wavelength, and
our XSTAR computation of the optical depth at the line core of each of these
lines when the plasma is at these two ionization states (this gives
a quantitative idea of the contribution to the complex).


\subsection{Case (ii)}

Nevertheless, after proving that the addition of a second      			%
component to the photoionized gas, is statistically required      		%
by the fits in both sets of data, we are able to support the     		%
identification of the feature as a Fe~{\sc xxv} line. 	      			%
The strongest evidence is that the high-velocity component    			%
is consistently fitted with an ionization parameter            			%
$\log\xi(2)\approx 3$. This is precisely where the peak of    			%
the ionization fraction of Fe$^{+24}$ is located. Therefore, we identify 	%
the absorption feature at $\approx 8$ keV as the complex Fe C1, in which	%
the major contributor is the K$_\mathrm{\alpha}$ line of Fe~{\sc xxv}.				%
The physical implications are summarized in Sect. \ref{discuss}.

\section{Discussion and Summary
\label{discuss}}

We have found several interesting aspects in the X-ray spectrum
of the QSO \apm.
The physical scenarios raised after the interpretation
of \cite{hasinger2002a} as a Fe edge, and of \cite{chartas2002a}
as relativistic ($\sim 0.2c-0.4c$) Fe~{\sc xxv} lines, have helped
to scrutinize more closely the X-ray spectrum of this object. 
If we attempt to explain the spectrum by
{\it only} investigating the full band $\sim 0.2-10$ k~e~V (observed frame),
we find it hard to justify which model represents
the data better in a statistical sense.
After including a local criterion to evaluate the fits in     	 %
the band $1-2$ k~e~V, we were able to converge to a model that   %
accounts for both low- and high- energy spectral bands.      	 %
Indeed, by modeling the spectrum with a photoionization model,
which gives a better (over Gaussian lines alone for instance) 
representation of edges and absorption lines
in the whole band $\sim 0.2-10$ keV, 
we are able to draw more physical information. 
                                                             	%
                                                              	%
                                                              	%

We discuss the implications of our results from Sect. \ref{photo}
in two parts:
(i) Assuming that the X-ray spectrum of \apm \sp can be modeled by
a {\tt single-absorber} photoionization model.
(ii) Assuming that the best representation of the
spectrum is a {\tt two-absorbers} model.

(i) {\it Changes in the physical and kinematical state of the absorber:} 
We see that the \xmm \sp data is represented by a highly-photoionized 
gas with column density $\sim 10^{23}$ \cmn, and
$\log(\xi)\sim 1.5$ at 0 \kms \sp in the frame of the quasar. Nevertheless,
by allowing iron to be higher than {\it solar} (model B), we obtain
a iron abundance best-fit value of
${\rm Fe}=5.04^{+0.44}_{-0.23}$, and the ionized gas column
no longer needs to be as high as $10^{23}$ \cmn, but $\sim 5\times 10^{22}$ \cmn
\sp instead, statistically improving the fit at 99\% of significance over model A
(with the same number of parameters than the absorbed power-law modified by an edge),
similar to the H02 interpretation.

But $\sim$ 2 weeks before
(in the frame of the quasar), the \chan \sp spectrum presented evidence of a
material outflowing $\sim 0.21c$, not only showing an absorption-line-like
feature at $\sim 8.1$ keV (rest-frame), but also pointing to absorption at lower
energies $\sim 0.8-2$ keV (rest-frame), coming from species of iron from
Fe~{\sc xviii} to Fe~{\sc xxiv}, and H- and He-like ions of Ne, Mg, Si, S, and O.
In this context, the \chan \sp data appears to be pointing to
(through model D):
(a) a change in the kinematical state of the absorbing material,
or more plausible, change in the projection of the velocity field on
our line of sight (which would yield to the same effect),
from $\sim 0.21c$ to $\sim $ 0 \kms \sp (since the \chan \sp 
observation was made first than the \xmm \sp observation); and
(b) a slight but noticeable change in the
ionization state of the gas, from $\log(\xi) \sim 2$ to $\log(\xi) \sim 1.5$
in a timescale of $\sim 2$ weeks in the frame of the quasar.
However, we find it physically hard to explain the deceleration shown
by this model, in addition to a change in the ionization state of the gas
without a significant change in the intrinsic luminosity of the source, 
and once we found the {\tt two-absorbers} model statistically superior,
we ruled out (i) and focused on (ii).

(ii) {\it Both absorbers have been there}:
The {\tt two-absorbers} model consists of: one absorber at     			%
$v=0$ \kms, $N_H(1)=7.34 _{-0.15} ^{+0.18}\times 10^{22}$ \cmn, 		%
$\log\xi(1)=1.50 _{-0.04} ^{+0.02}$, and a second absorber at    		%
$v=0.18c$, $N_H(2)=5.49 _{-0.54} ^{+0.71}\times 10^{22}$ \cmn,			%
$\log\xi(2)=3.00 _{-0.01} ^{+0.04}$ 						%
(parameters coming from the simultaneous best-fit).				%
The chemical composition							%
of both is the same, {\it solar} in all the elements (see model 		%
composition in Sect. \ref{photo}) except in the abundance of iron			%
${\rm Fe}=2.85 _{-0.10} ^{+0.10}$.						%
It is worth mentioning that from the separated fits, the two observations show a change
in the power-law of $\approx 10$\% and also show changes in the physical parameters.
However, they (individually) change little (i.e, no large difference within errors);
$\log\xi(1)$, from $\approx$ 1.2 to 1.5 ($\approx$ 30\%);
$N_H(1)\times 10^{22}$ \cmn, from $\approx$ 5.7 to 6.1 ($\approx$ 10\%);
$\log\xi(2)$, from $\approx$ 3 to 3.1 ($\approx$ 3\%);
$N_H(2)\times 10^{22}$ \cmn, from $\approx$ 4.8 to 2.7 ($\approx$ 70\%, but see errors).
The computation of $\chi^2$, on the other hand, is very sensitive to the continuum level,
and small changes in the power-law are easily detected by the minimization-$\chi^2$
routine. {\it We do not think this is a fundamental physical change of state} between
observations.
At the moment these data do not allow us to discriminate whether
the changes seen in the physical parameters are significant (within $\approx 2\sigma$) or not.
Apparently, apart from the power-law component, both observations can be represented
by the same two-absorbers model.

We have verified that the overabundance of Fe can be established		%
respect to the abundance of oxygen (since it is the Fe/O ratio, which is	%
cosmologically relevant), by comparing the observed spectra			%
with the synthetic theoretical spectra at low-energies, if we increase		%
the oxygen abundance. The ratio Fe/O must be $\approx 3$ in order to obtain	%
acceptable fits in the low-energy band.						%
\footnote{We verified that the data is sensitive to an increase of oxygen abundance.
We take the best-fit two-absorbers model and compute models with the oxygen abundance
at 1.5, 3 and 5 $\times$ \emph{solar} oxygen abundance. The fit get worse with
$\chi^2({\rm O=1.5})=304$, $\chi^2({\rm O=3})=333$ and $\chi^2({\rm O=5})=388$.
The conclusions are: 1) the data is sensitive to the Fe/O ratio. 2) This ratio must
be $\approx 3 \pm 1$, in order to produce acceptable fits 
(and produce good description of the low-energy band).}
This was first noticed by \cite{hasinger2002a}, and its        			%
cosmological implications are discussed in \cite{komossa2003a},			%
based on chemical evolution studies of \cite{hamann1993a}.			%

%
This multi-component photoionization approach has been       			%
successfully applied to other AGNs in different bands (X-ray and UV): 		%
NGC 3783 \citep{netzer2003a,krongold2003a,gabel2005a};				%
Mrk 279 \citep{fields2007a,costantini2007a};					%
NGC 4593 \citep{steenbrugge2003b};						%
NGC 985 \citep{krongold2005a};							%
\mr \sp \citep{kaspi2004a};							%
NGC 4051 \citep{kraemer2006a,armentrout2007a,krongold2007a}, 			%
a survey of a sample of 15 Type I AGNs can be found in \cite{mckernan2007a}.	%
We will treat, this multi-component photoionization approach, as the best 	%
solution for our problem. 							%
However, note that a key difference with some of these cited papers is that
the solution for \apm \sp requires two unrelated absorbers at very different
velocities.
                                                              %
                                                              %
                                                              %

A plausible framework in which to place these observational clues is in
accretion disk wind models
\citep{murray1995a,elvis2000a,proga2000a,proga2007a}. Recent high-resolution
spectroscopy of \apm, shows UV BALs in a wide range of
velocities ($5\,000 \lesssim v \lesssim 12\,000$ \kms) and ionization
degrees (ions like C~{\sc vi}, O~{\sc vi}, N~{\sc v} and Si~{\sc iv}), 
implying that the outflow may be
composed of multiple components, with velocity; density and/or
ionization parameter gradients \citep{srianand2000a}.
In particular, the model proposed by \cite{elvis2000a}
predicts that the UV BAL is formed in the conical section
of a funnel-shaped flow ranging velocities
$10\,000 \lesssim v \lesssim 60\,000$ \kms, (or up to $\sim 0.2c$
in our context). Furthermore, the model is very specific
as to the location and physical properties of the
X-ray absorbing material. The warm highly-ionized medium
has column densities of $\sim 10^{22.5}$ \cmn, and
temperatures of $\sim 10^{6}$ K, right pressure, and
ionization parameter to ensure pressure equilibrium with
the BELR clouds.
Our XSTAR-based photoionization model
estimates that the high-ionization, high-outflow velocity component (HV component),
$\log(\xi)\approx 3$ has a temperature $T\sim$ few times $10^6$ K,
and the best-fit column density
$N_H(2) \sim 10^{22.5}$ \cmn,
nicely consistent with the main physical properties of the warm highly-ionized medium.
The low-ionization rest-frame velocity component (RV component)
with 
$\log(\xi)\approx 1.5$ has a temperature $T\sim$ $10^5$ K,
in accordance with the temperature of $\sim$ few times $10^5$ K
found by \cite{kaastra1995a} in NGC 5548, and the degree of
ionization is consistent with $\log(U) \sim 0.3$
reported by \cite{mathur1995a} for the same object, where
a test to unify UV/X-ray absorbers in Seyfert galaxies
was successfully applied.

Recalling the definition of the ionization parameter $\xi=L_{\rm ion}/(nR^2)$,
where $L_{\rm ion}$ is ionizing luminosity, $n$ is the gas density,
and $R$ is the location of the absorber, we are able to compute
the product of two un-observable quantities $(nR^2)$
from the observable $\xi$ and $L_{\rm ion}$. We measured an intrinsic
X-ray luminosity
\footnote{This is the unabsorbed intrinsic $(1-50)$ keV rest-frame luminosity
of the source. From the \xmm \sp data 
$L_{\rm x}(\xmm)=1.07(\pm0.03)\times 10^{47}k^{-1}$ \ergs,
from the \chan \sp data
$L_{\rm x}(\chan)=1.17(\pm0.05)\times 10^{47}k^{-1}$ \ergs.
We are taking the average of both.}
of $L_{\rm x}=1.12(\pm0.04)\times 10^{47}k^{-1}$ \ergs, $k$ being the lensing
magnification factor. With this we have:
$(nR^2)$[HV]$=1.12(\pm 0.11)\times10^{44}k^{-1}$ cm$^{-1}$ and 
$(nR^2)$[RV]$=3.54(\pm 0.11)\times10^{45}k^{-1}$ cm$^{-1}$. This is the
first time this quantity is measured for this BAL QSO 
(since it is the first time $\xi$ is properly constrained).
Despite
we still are not able to break the natural degeneracy between $n$ 
and $R$ (due to lack of variability),
it allows us to present order-of-magnitude estimates on the
mass-loss rate and energy budgets of the system.
Assuming the HV forms part of the high-ionization BAL flows, which reaches
velocities up to $\sim 0.2c$ \citep[see][ for a recent review of high
velocity (HV) outflow in quasars]{rodriguez2007a}, and that the
X-ray absorber in \apm \sp is part of the conical section of the funnel-shaped
flow proposed by \cite{elvis2000a} (although this estimation is independent of
this specific geometry and can be equally applied to a spherical shell of gas
for instance, without significatively changing the final conclusions), 
which forms a shell with covering factor
$f_{\rm cov}$
moving at a velocity $v$, we estimate the mass-loss
rate for the system:
\begin{equation}
\dot{M}_{w}=
4\pi \mu_H (nR^2) vf_{\rm cov},
\end{equation}
where $\mu_H$ is the mean mass per H atom.
Using $(nR^2)$[HV]$=1.12\times10^{44}k^{-1}$ cm$^{-1}$,
an outflow velocity $v=0.18c$ and assuming a covering factor of 0.01,
we have a mass-loss rate of
$\dot{M}_{w}\approx 2873k^{-1}$ $M_{\odot}~{\rm yr}^{-1}$. The total
kinetic power released by the wind in \apm \sp is
$\dot{M}_{w}v^2/2=2.6\times 10^{48}k^{-1}$ \ergs.
                                                              %
                                                              %
                                                              %
                                                              %
                                                              %
                                                              %
The accretion rate is related with the accretion luminosity with:
\begin{equation}
\dot{M}_{\rm acc} \approx 17 \frac{L_{\rm bol}}{10^{47}~{\rm ergs~s^{-1}}}\frac{0.1}{\eta}
M_{\odot}~{\rm yr}^{-1},
\end{equation}
where $L_{\rm bol}$ is the bolometric luminosity,
$\eta$ is the accretion efficiency (assumed to be $\sim 0.1$),
and $c$ is the speed of light. 
With our measured $L_{\rm x}$ we are able to estimate $L_{\rm bol}=\beta L_{\rm x}$.
We adopt a bolometric correction factor $\beta=20$,
based on the mean AGN SED of \cite{elvis1994a}.
Explicitly, in terms of the lensing magnification factor, 
$\dot{M}_{\rm acc} \approx 340k^{-1}$~ $M_{\odot}~{\rm yr}^{-1}$.
This would lead to the conclusion that the central black hole in
\apm \sp {\it accretes $\sim 1$ order of magnitude less matter than the estimated
mass-loss rate}.
Recently, from high-resolution \chan \sp and \xmm \sp spectroscopic studies
at least three systems show the property of having mass outflow rates
$\sim 1$ order of magnitude larger than the accretion rates (at $\sim $ 10\% efficiency):
the micro-quasar GRS 1915+105 \citep{lee2002a}, 
the Seyfert 1 galaxy MCG-6-30-15 \citep{turner2004a,sako2003a,lee2002b},
and the Seyfert 2 galaxy IRAS 18325-5926 \citep{lee2005a}.
However, we are aware of two major sources of uncertainties in these estimations.
Due to its location, the uncertainty in $f_{\rm cov}$ is large because the
exact geometry in the base of the outflow is not known. On the other hand,
the bolometric correction factor could be larger than the one used here,
locating $\dot{M}_{\rm acc}$ and $\dot{M}_w$ in the same order of magnitude.
Either we detect the outflow in a very small period of activity 
(which would be very fortuitous) or the outflow
plays a major role in the process of sending
matter back to the interstellar medium \citep[e.g.,][]{crenshaw2003a,lee2005a,krongold2007a}.
If we assume that the mass of the
black hole $M_{\rm BH}\approx 2\times 10^{10}$ $M_{\odot}$,
\citep[as estimated by][]{shields2006a},
the ratio $L_{\rm bol}/L_{\rm edd}\approx 0.8k^{-1}$
(where $L_{\rm edd}$ is the Eddington luminosity).
This number is in accordance with the general trend of low
$L_{\rm bol}/L_{\rm edd}$ ratio for BALQSOs reported by \cite{boroson2002a}.
With these estimations, we can derive the launching radius $r_{\rm launch}$
for a radiatively-driven wind:
\begin{equation}
r_{\rm launch}\approx \frac{2GM_{\rm BH}}{v^2}
\left( \Gamma_f \frac{L_{\rm UV}}{L_{\rm edd}}-1\right).
\end{equation}

Here, we have used the relation $v~vs~r$ for a radiatively-driven
steady wind  
and assumed that at very large radii the flow has reached a
terminal velocity $v$. Assuming that the wind is driven by a UV
luminosity $L_{\rm UV}\sim 4\times 10^{46}$ \ergs, with a force
multiplier $\Gamma_f\sim 100$ \citep[taken from][]{laor2002a},
and using a flowing velocity $v=0.18c$, we find
$r_{\rm launch}\approx 7\times 10^{16}$ cm.
On the other hand, if we use the maximum observed velocity of
the UV absorber (through the C~{\sc vi} line), for this quasar
of $v=0.041c$, we find $r_{\rm launch}\approx 1\times 10^{18}$ cm.
One possible explanation for the difference in the two distances,
is that the UV absorber is at a lower ionization state, lying in
the low ionization BAL region of the funnel-shaped flowing
\citep[for visual help see Fig. 5 of][]{elvis2000a}, 
and more specifically in the large radii part in which the UV absorber 
has reached its highest (terminal) velocity.
A rigorous study of the hydrodynamical properties of the X-rays/UV
absorbers in AGN is beyond the scope of this paper. However,
the simplistic approach used in this analysis could be useful as a
framework for future works on the study of the central region of this
quasar.

To summarize, the absorbing gas in \apm \sp can be represented by a {\tt two-absorbers} model with
column densities $N_H(1)\approx 7\times 10^{22}$ \cmn, $N_H(2)\approx 6\times 10^{22}$ \cmn,
ionization parameters 
$\log\xi(1)\approx 1.5$ and $\log\xi(2)\approx 3$, with one of them (the HV component) outflowing 
at $v\approx 0.18(\pm 0.01)c$,
carrying large amount of gas out of the system.
The feature at $\sim 8$ keV (rest-frame) is fully predicted and reproduced by
our photoionization model, to be a complex of Fe lines coming from high state
of ionization, in which the main contributor is the Fe~{\sc xxv} $\lambda 1.85$ \AA,
improving the characterization of the kinematics and the quantitative evolution
analysis of this high $z$ quasar.
We confirm evidence for an 
overabundance of Fe/O, from the \xmm \sp observation of this quasar
\citep[previously inferred and discussed in][ based on calcutations of Hamann \& Ferland 1993]{hasinger2002a,komossa2003a}.
The analysis is made for the first time on the \chan \sp observation of
\apm, implying that both absorbers require Fe/O supersolar, placing similar
constraints on models as before, and additionally shows that both independent absorbers
have a similar chemical history.

\begin{acknowledgements}

This work is based on observations obtained with \xmm,
an ESA science mission with instruments and contributions
directly funded by ESA Member States and the US
(NASA). In Germany, the \xmm \sp project is supported by the
Bundesministerium f\"ur Wirtschaft und Technologie/Deutsches Zentrum
f\"ur Luft- und Raumfahrt (BMWI/DLR, FKZ 50 OX 0001) and the Max-Planck
Society. It is also based on an observation obtained with the Chandra X-ray
telescope (a NASA mission).
The author wants to thank Stefanie Komossa for her important contributions,
for suggesting the topic and ongoing discussions
throughout the work. He also is grateful to G\"unther Hasinger
for suggesting improvements in the final presentation of the work.
This work was supported through a postdoctoral position
at the
Max-Planck-Institut f\"{u}r extraterrestrische Physik
(MPE, Garching-Germany),
with partial contribution from the DFG Leibniz Prize 
(FKZ HA 1850/28$-$1).

\end{acknowledgements}



\end{document}